\documentstyle{l-aa}
\input epsf.sty
\def\pa{\parallel}
\def\pe{\perp}

\def\eps{\varepsilon}

\def\s{\sigma}
\def\o{\omega}

\def\a{\alpha}
\def\b{\beta}
\def\d{\delta}
\def\D{\Delta}

\def\g{\gamma}
\def\th{\theta}
\def\ri{\right}
\def\le{\left}

\def\bc{\bigg}
\def\bd{\Bigg} 

\begin{document}

\thesaurus{06 (02.18.5; 08.16.7)}

\title{Orthogonal polarization mode  phenomenon in pulsars}

\author{R. T. Gangadhara}

\institute{Max--Planck--Institut f\"ur Radioastronomie, Auf dem H\"ugel 69, D--53121 Bonn, 
Germany\\
ganga@mpifr-bonn.mpg.de}

\date{Received 14 October 1996/ Accepted 2 June 1997}

\maketitle

\begin{abstract}

    We consider the polarization properties of radiation emitted by relativistic charged particles
while moving along the curved magnetic field lines in a pulsar manetosphere. We propose that the 
radiation emitted by positrons and electrons 
while moving along the curved magnetic field lines is
orthogonally polarized. The polarization angle of each orthogonal polarization mode 
is well described by the rotating vector model. However, the polarization angle swings observed
in micropulses and subpulses are often found to be contrary to the 
rotating vector model, 
and such swings are expected to arise due to the coherent superposition of 
orthogonal polarization modes. 
\par
   As an application of our model, we discuss the  polarization of pulsar PSR~B0950+08 at the
frequency 1.71~GHz. Data on individual pulses, obtained using the 100-m Effelsberg radiotelescope,
was statistically analysed and the results are presented as probability of occurrence gray-scale
plots. We find that pulsars emit radiation mainly in the form of two independent 
orthogonal modes. 
It seems, they exist at all pulse longitudes but at some longitudes one mode dominates over the other. 
The polarization angle gray-plot indicates two most favoured  angles separated by approximately $90^o$ 
at each pulse longitude. The depolarization is mainly caused by the incoherent superposition of 
orthogonal modes. 
\par
   We infer from this study that pulsar radiation consists of two major elliptically 
polarized electromagnetic waves with orthogonal polarization angles.
Our model predicts that such waves could be radiated by the positrons and the electrons
accelerated along the curved magnetic field lines. The sense of circular polarization
of these modes depend upon from which side of particle trajectory the radiation is received. 

\keywords{Pulsars: individual: PSR~B0950+08--radiation mechanism: non-thermal}

\end{abstract}

\section{Introduction}

Pulsar radio emission and further propagation effects in the magnetosphere are not well understood.
This situation 
is partly due to the difficulty in understanding particle acceleration and current flow in pulsars superstrong 
magnetic field. From the observation point of view curvature radiation seems to be most attractive 
among all the proposed radio emission mechanisms (Michel 1991; M$\acute{\rm e}$sz$\acute{\rm a}$ros 1992;
Xilouris et al. 1994). However the coherent curvature emission by bunches has been often criticized (e.g. Melrose
1981). Curvature radiation is quite similar to synchrotron radiation, with the only difference that
the role of Larmor radius is played here by the radius of curvature of magnetic field lines.
Michel (1987a) has indicated that curvature radiation has significant circular polarization when viewed at an 
angle to the plane of a magnetic field line, and reverses sense when viewed
from the other side. 
\par
The curvature radiation model has been developed by assuming that particles follow the curved field lines 
(Sturrock 1971; Ruderman \& Sutherland 1975). The theory of synchrotron radiation
is developed by assuming uniform and straight field lines while the curvature radiation
is developed by ignoring the spiral motion of the particles.  The most often cited reason for
ignoring spiral motion is that particles will instantly lose the energy associated with perpendicular
component of motion by synchrotron radiation, cascading down to zero free energy. This is true near the polar cap
where the field lines are straight. But when the particles move into the curved region of field lines
they recover the perpendicular component of momentum as the field lines curve off from the direction
of particle velocity. Hence the radiation emitted by charged particles while moving along the curved
magnetic field lines cannot be described by  considering either synchrotron or simple curvature radiation 
separately. Gil \& Snakowski (1990) have attempted to examine the polarization properties of 
curvature radiation but they have also not considered the role of magnetic force. 
\par
One of the most fascinating features of pulsar radiation is the occurrence
of orthogonal polarization modes (OPM), i.e., the two electromagnetic waves with orthogonal 
electric vectors. This phenomenon has become difficult to explain by emission models based on the
simple curvature emission, as it cannot specify the two preferred polarization angles.
There are other models based on the
propagation effects but one would expect such effects to be strongly dependent on frequency and require
special viewing angles (Michel 1991), whereas observations indicate the phenomenon is broad band.
The interpretation based on the geometrical effects (Michel 1987b) indicates that two modes could be 
due to the overlap of radiation from two distinct emission regions in the magnetosphere. However,
it is not clear why the two separated sources have nearly orthogonal polarization.
\par
       Pulsars have been noted for their highly polarized radiation. To understand the phenomenology of 
pulsar polarization, many attempts have been made to fit the average polarization angle swing within the
context of the rotating vector model (RVM) (Radhakrishnan \& Cooke 1969). However many pulsars do not
fit with this interpretation (Manchester 1971; Rankin et~al. 1974) and such discrepancies have 
been attributed to the occurrence of OPM (Backer et~al. 1976; Gil \& Lyne 1995). The relative 
strengths of OPM vary with pulse longitude, thus causing discontinuities in the average pulse 
polarization angle swing (Cordes \& Hankins 1977). The switching between OPM mostly but 
not always occurs on the boundaries of micropulses and subpulses (Cordes \& Hankins 1977), and may be 
intrinsic to the emission process (Manchester et~al. 1975; Cordes et~al. 1978).
The single-pulse studies (Stinebring et~al. 1984a,b) show that OPM
overlap, and in the vicinity of pulse longitude where jump occurs some pulses prefer to have one polarization
and some have other polarization. All these fine details are lost when the pulses are averaged. 
The depolarization in average pulse occurs due to the superposition of OPM 
(Lyne et~al. 1971; Manchester et~al. 1975; Stinebring et~al. 1984a). 
\par
       Cordes et~al. (1978) have reported the existence of OPM with opposite senses of circular 
polarization in pulsar PSR~B2020+28 at 430 MHz. They found no evidence for a threshold
intensity in the occurrence of OPM. The peak subpulse emission can be in either OPM, and sometimes 
transitions between modes do not occur on the edges of subpulses. Therefore, the occurrence of OPM
may be stochastic and perhaps indicating the untenability of a geometric interpretation of the 
transitions as aspects of an angular beam of radiation. 
\par
       The motion of a charged particle along a rotating magnetic field line is discussed
by Gangadhara (1995; 1996a) and in the case of a curved magnetic field line (Gangadhara 1996b), and
the studies indicated that OPM may be produced by positrons and electrons, as they have got 
opposite senses of gyration.
In this paper, we show that the relativistic positrons and electrons can produce OPM because while moving
along a curved magnetic field line their accelerations  become inclined with respect to 
the radius of curvature of a field line. We propose the coherent superposition of OPM as a physical
explanation for the polarization angle swings observed in micropulses and subpulses.
In Sect.~2 we discuss the forces acting on a positron and an electron while moving along a curved 
magnetic field line. The radiation fields due to a positron and an electron are derived
in Sect.~3. The polarization properties of radiation fields are computed in Sect.~4. Finally, in
Sect.~5, we present a series of gray-plots describing the polarization states of OPM from PSR~B0950+08, and 
discuss the possible explanation based on our model.

\section{Motion of relativistic charged particles along curved magnetic field lines}

     In the physics of pulsar magnetosphere, a very important question is how are the charged particles 
ejected from the surface of a neutron star, and  it has been discussed by many authors
(e.g. Herring \& Nichols 1949; Good \& M\"uller 1956; Ruderman \& Sutherland 1975; Kundt \& Schaaf 1993).
If particles are produced with the velocity having a component perpendicular to the magnetic field 
they quickly radiate away the energy associated with that component of motion by synchrotron radiation. Hence 
the particle gyration becomes almost absent near the polar cap. Once the particles radiate away 
the energy associated with the perpendicular component of velocity, they are free to move along the 
field lines as long as they are straight.
But when the particles enter the curved region of field lines, they recover perpendicular component of 
velocity at the expense of parallel component (Gangadhara 1996b) as the field line curves off from the 
direction of particle motion (Fig.~1). This phenomenon must be true for both primary as
well as secondary particles.

\begin{figure}
\vskip -13.5 truecm \epsfysize=10. truecm
\centerline{\hskip 1.5 truecm\epsffile[00 350 450 790]{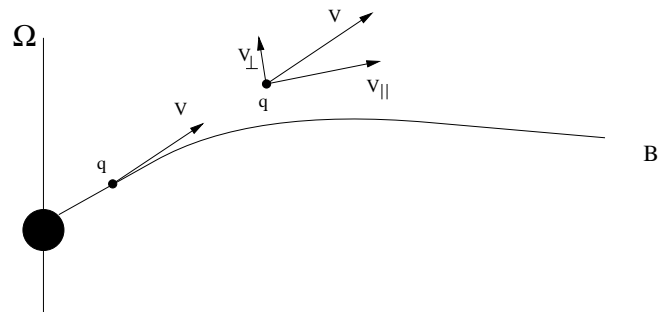}}
\vskip 8.0 truecm
\caption{Motion of a particle with charge $q$ and velocity ${\bf v}$ along a curved field line ${\bf B}$.}
\end{figure}

    The motion of a particle along a curved magnetic field line is governed mainly by the magnetic Lorentz force 
$\vec F_{\rm B}$ and the centrifugal (inertial) force $\vec F_{\rm c}.$ Figure~2 shows the directions of these
forces when a positron and an electron are in motion along a curved magnetic field line. The net force acting on a particle is given by
\begin{equation}
\vec F_{\rm i}=\vec F_{\rm ci}+\vec F_{\rm Bi},
\end{equation}
where i $=$ p for a positron and e for an electron. The magnetic force is
\begin{equation}
\vec F_{\rm Bi}=\frac{q_{\rm i}}{c}\vec v_{\rm \pe i}\times\vec B
\end{equation}
and the centrifugal force is
\begin{equation}
\vec F_{\rm ci}=\frac{\g_{\rm i} m_{\rm i} v_{\rm \pa i}^2}{\rho^2}\vec \rho,
\end{equation}
where $m_{\rm i}$ and $q_{\rm i}$ are the mass and charge, $\g_{\rm i}$ is the relativistic Lorentz factor, $v_{\rm \pa i}$
and $v_{\rm \pe i}$ are the components of velocity with respect to the magnetic field B, respectively. Here $\vec \rho$ 
is the radius of curvature of a magnetic field line.

\begin{figure}
\vskip -13. truecm \epsfysize=10. truecm
\centerline{\hskip 1.5 truecm\epsffile[00 350 450 790]{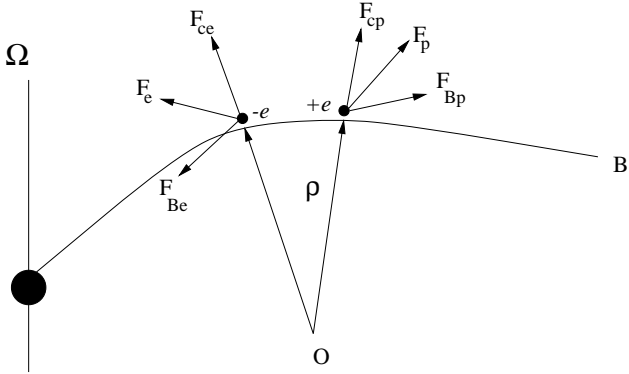}}
\vskip 8.0 truecm
\caption{The forces ${\bf F}_{\rm Bp}$ and ${\bf F}_{\rm cp}$ act on  a positron,
and ${\bf F}_{\rm Be}$ and ${\bf F}_{\rm ce}$ act on an electron while they are in motion along a curved field line
${\bf B}.$ Here ${\bf \rho}$ is the radius of curvature of field line.}
\end{figure}

    Initially, magnetic and centrifugal forces act in the directions perpendicular and parallel to the plane 
of magnetic field line, respectively. The motion of particles along curved field lines
is discussed by Jackson (1976) and Gangadhara (1996b). The gyration motion 
(spiral motion) is considered as the zero--order-motion, and the motion that arises due to the curvature of field 
lines as the first-order motion. In the absence of initial perpendicular component of velocity, particles cannot 
gyrate around the curved magnetic field lines. The reason is that as particle attempts to spiral, field 
line curves off to the side and hence the centrifugal force prevents it from spiraling (Gangadhara 1995; 1996a). 
Also, it is observationally known that no generalized Faraday rotation is evident in pulsar magnetospheres
(Cordes 1983; Lyne \& Smith 1990). There is {\it no central force} at O, and 
the magnetic force plays the role of centripetal force to keep the particle on 
track. But the
peculiarity of magnetic force is that it acts in the directions perpendicular to the magnetic field and not always in the direction of radius of curvature. 
{\it Since $\vert\vec F_{\rm Bi}\vert\approx\vert\vec F_{\rm ci}\vert,$ the net forces 
$\vec F_{\rm p}$ and $\vec F_{\rm e}$ are inclined through, respectively, $\sim 45^o$ and 
$-45^o$ with respect to the plane of magnetic field line.} Therefore, the net 
accelerations of positrons and electrons become orthogonal in the curved magnetic
field lines of pulsar magnetosphere. The detailed computation of particle dynamics,
including radiation reaction, is under consideration and published elsewhere.

\section{Radiation fields of positrons and electrons accelerated in curved magnetic field lines}

   Consider a relativistic particle with charge q moving along a curved trajectory C (space
curve) in the xyz--coordinate system (Fig.~3). Let $\th$ be the angle between 
particle position $\vec r$ and yz-plane. A distant observer at P receives radiation at an angle $\phi_{\rm p}$ from the plane of particle orbit. The electric field of radiation at the observation point
is given by (Jackson 1976)
\begin{equation}
\vec E(\vec r, t)=\frac{q}{c}\left[\frac{\hat{k}\times[(\hat{k}-\vec
\b)\times\dot{\vec \b}]}{S\, \s^3}\right ]_{\rm ret},
\end{equation}
where $\s=1-\vec\b\cdot\hat k.$ The distance from radiating region to the observer 
is $S,$ the propagation vector is $\hat k$, and the velocity and
acceleration of particle are $\vec \b=\vec v/c$ and $\dot{\vec\b}.$
\par
The radiation emitted by a relativistic charged particle has a broad spectrum.
The range of frequency spectrum is estimated by
taking Fourier transformation of electric field of radiation:
\begin{equation}
\vec E(\o)=\frac{1}{\sqrt{2\pi}}\int\limits_{-\infty}^{+\infty}\vec E(t) {\rm e}^{i\o t} dt. 
\end{equation}
In Eq.~(4), ret means evaluated at the retarded time $t'+\frac{S(t')}{c}=t.$ By changing the variable of integration from t to $t',$ we obtain
\begin{equation}\vec E(\o)=\frac{1}{\sqrt{2\pi}}\frac{q}{c}\!\!\int\limits_{-\infty}^{+\infty}
\frac{\hat{k}\times[(\hat{k}-\vec \b)\times\dot {\vec \b}]}{S\, \s^2}
{\rm e}^{i\o\{t'+S(t')/c\}} dt', 
\end{equation}
where we have used $dt=\s dt'.$
\par
   When the observation point is far away from the region of space where the
acceleration occurs, the propagation vector $\hat k$ can be taken constant in
time. Furthermore the distance $S(t')$ can be approximated as 
\begin{equation}S(t')\approx S_{\rm o}-\hat{k}.\vec r(t'),
\end{equation}
where $S_{\rm o}$ is the distance between the origin O and the observation point P, and
$\vec r(t')$ is the position of particle relative to O.
Then Eq.~(6) becomes
\begin{equation}\vec E(\o)\approx \frac{q\, {\rm e}^{i\o S_{\rm o}/c}} {\sqrt{2\pi} S_{\rm o}c} \!
\int\limits_{-\infty}^{+\infty}
\frac{\hat{k}\times[(\hat{k}-\vec \b)\times \dot{\vec \b}]}{\s^2}
{\rm e}^{i\o\{t-\hat k.\vec r/c\}} dt, 
\end{equation}
where the primes on the time variable have been omitted for brevity. The
integrand in Eq.~(8), excluding exponential, is a perfect differential, therefore,
we can integrate by parts, and obtain 
\begin{equation}\vec E(\o)=-i \frac{q\o\, {\rm e}^{i\o S_{\rm o}/c}} {\sqrt{2\pi} S_{\rm o} c} 
\!\!\int\limits_{-\infty}^{+\infty}\!\! \hat{k}\times(\hat{k}\times \vec \b)
{\rm e}^{i\o\{t-\hat k.\vec r/c\}} dt. 
\end{equation}
The polarization of emitted radiation can be estimated for a specified motion
with known $\vec r(t)$ and $\vec \b.$

\begin{figure}
\vskip -10.5 truecm \epsfysize=8.5 truecm
\centerline{\hskip 0.0 truecm\epsffile[00 350 450 790]{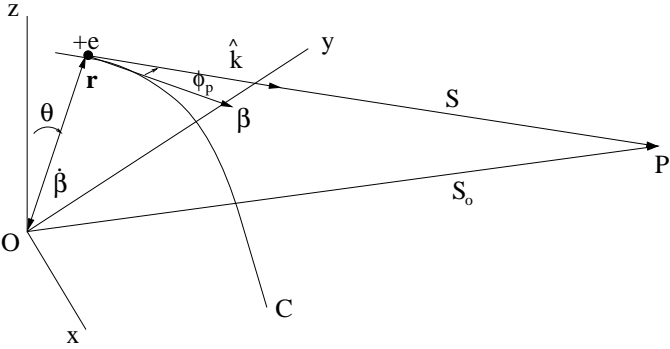}}
\vskip 7.0 truecm
\caption{The geometry used to describe the radiation emitted by a positron while moving
   along a curved trajectory C. Here $\th$ is the angle between yz-plane and 
   ${\bf r},$ and $\phi_{\rm p}$ is the angle between ${\bf \beta}$ and $\hat k.$}
\end{figure}

\subsection{Radiation field due to a positron}

    Since the duration of pulse $\D t\sim R/c\g$ is very short, it is
necessary to know the position $\vec r(t)$ and velocity $\vec \b$ of particle
over only small arc of the trajectory whose tangents are in the general
direction of observation. Therefore, for a positron moving under the action of
magnetic and centrifugal forces, we take:
\begin{equation}\vec r(t)=R\le(\sin\th,\, \frac{\cos\th}{\sqrt{2}},\, \frac{\cos\th}{\sqrt{2}}\ri), 
\end{equation}
\begin{equation}\vec \b(t)=\b\le(\cos\th,\, -\frac{\sin\th}{\sqrt{2}},\, 
-\frac{\sin\th}{\sqrt{2}}\ri). 
\end{equation}
Since the integral in Eq.~(9) has to be computed over the path of 
particle, the unit vector $\hat k$ can be chosen, without loss of 
generality, to lie in a plane which is parallel to the xy-plane:
\begin{equation}\hat k=(\cos\phi_{\rm p},\, \sin\phi_{\rm p},\, 0), 
\end{equation}
where $\phi_{\rm p}$ is the angle made by $\hat k$ with respect to the positron
velocity.
Using Eqs.~(11) and (12), the vector part of the integrand in Eq.~(9)
can be written as
\begin{eqnarray}
&&\!\!\!\!\!\!\!
\hat{k}\times(\hat{k}\times \vec \b)= \b\! \bc[\frac{\sin\th}{\sqrt{2}} \hat\eps_\pa+\!\!\bc(\sin\phi_{\rm p} \cos\th +
\frac{ \cos\phi_{\rm p} \sin\th }{\sqrt{2}}\bc) \hat\eps_\pe\! \bc] \nonumber\\
&&
\end{eqnarray}
where $\hat \eps_\pa=\hat z$ is a unit vector in the direction of z--axis, and 
$\hat\eps_\pe=- \sin\phi_{\rm p} \hat x + \cos\phi_{\rm p} \hat y =\hat\eps_\pa\times\hat k$ is a unit
vector which is orthogonal to both $\hat\eps_\pa$ and $\hat k.$
\par
Using Eqs.~(10) and~(12), the argument of exponential in Eq.~(9) can be
written as
\begin{eqnarray}
&&\!\!\!\!\!\!\! \o\left(t-\frac{\hat k.\vec r}{c}\right)=\o\left[t-\frac{R}{c}\left(
\cos\phi_{\rm p}\sin\th+\frac{\sin\phi_{\rm p}\cos\th}{\sqrt{2}}\right)\right]. 
\nonumber \\
&& 
\end{eqnarray}
Since $R$ is close to the radius of curvature of particle trajectory, $\th$ can be replaced by $c \b t/R.$ 
Due to the relativistic beaming only for small values of $\phi_{\rm p}$ will there be an appreciable radiation,
and therefore the duration of pulse $\D t\sim R/c\g$ becomes very short. Hence the arguments of sine and cosine 
functions in Eqs.~(13) and (14) are of the order of $1/\g.$ So, we obtain 
\begin{eqnarray}
&&\!\!\!\!\!\!\!
\hat k\!\times\!(\hat k\times\vec \b)\approx\b\!\bc[\frac{\b c}{\sqrt{2}R}t
\hat\eps_\pa +\!\!\bc\{\frac{\b c }{\sqrt{2}R}t+
\phi_{\rm p}\!\le(1-\frac{c^2\b^2}{2R^2}t^2\ri)\!\!\bc\}
\hat\eps_\pe\!\bc ] \nonumber\\
&& 
\end{eqnarray}
and
\begin{equation}\o\left(
t-\frac{\hat k.\vec r}{c}\right)\approx\frac{\o}{2}\le[\le(\frac{1}{\g^2}
+\phi^2_{\rm p}\ri) t+\frac{c^2}{3R^2}t^3\ri],
\end{equation}
where $\b$ has been set equal to unity at wherever possible, and neglected all
those terms which are of the order of $1/\g^2$ times those kept.
\par
The components of electric field (Eq.~9) in the direction of unit vectors $\hat\eps_\pa$
and $\hat \eps_\pe$ are:
\begin{eqnarray}
E_\pa(\o)=-i\frac{q\b^2\o}{2\sqrt{\pi}S_{\rm o} R} {\rm e}^{i\o S_{\rm o}/c}\!\!
\int\limits_{-\infty}^{+\infty}\!\! t\,\exp &&\!\!\!\!\!\bc[i\frac{\o}{2}\bc\{\bc(
\frac{1}{\g^2}+\phi^2_{\rm p}\bc)t+\nonumber\\ &&\quad\frac{c^2}{3R^2}t^3\bc\}
\bc]dt, 
\end{eqnarray}
\begin{eqnarray}
&&\!\!\!\!\!\! E_\pe(\o)=-i \frac{q\b\o} {\sqrt{2\pi}S_{\rm o} c} {\rm e}^{i\o S_{\rm o}/c}
\int\limits_{-\infty}^{+\infty}\bc\{\frac{c\b}{\sqrt{2}R}t+\phi_{\rm p}\bc(1-
\nonumber\\
&&\,\,\,\frac{c^2\b^2} {2R^2}t^2\bc)\bc\}
\exp\!\left[i\frac{\o}{2}\left\{\left(\frac{1}{\g^2}+
\phi^2_{\rm p}\right)t+\frac{c^2}{3R^2}t^3\ri\}\ri]dt. 
\end{eqnarray}
To reduce the integrals into some known forms, we change the variable to
\begin{equation}y=\le(\frac{1}{\g^2}+\phi^2_{\rm p}\ri)^{-1/2}\frac{ct}{R} 
\end{equation}
and introduce a parameter
\begin{equation}\xi_{\rm p}=\frac{\o R}{3c}\le(\frac{1}{\g^2}+\phi^2_{\rm p}\ri)^{3/2}. 
\end{equation}
Therefore, the integrals are 
\begin{eqnarray}
I_{\rm 1} & = & \frac{R}{c}\le(\frac{1}{\g^2}+\phi^2_{\rm p}\ri)^{1/2}\!\int\limits_{-\infty}^{+\infty}\!\!\!
	\exp\le\{i\frac{3}{2}\xi_{\rm p}\le(y+\frac{y^3}{3}\ri)\ri\}dy,\\
I_{\rm 2} & = & \frac{R^2}{c^2}\le(\frac{1}{\g^2}+\phi^2_{\rm p}\ri)\!\!\int\limits_{-\infty}^{+\infty}\!\! y\,
	\exp\le\{i\frac{3}{2}\xi_{\rm p}\le(y+\frac{y^3}{3}\ri)\ri\}dy,\\
I_{\rm 3} & = & \frac{R^3}{c^3}\le(\frac{1}{\g^2}+\phi^2_{\rm p}\ri)^{3/2}\int\limits_{-\infty}^{+\infty}y^2
	\exp\le\{i\frac{3}{2}\xi_{\rm p}\le(y+\frac{y^3}{3}\ri)\ri\}dy.\nonumber\\
&  & 
\end{eqnarray}
We can identify these integrals with Airy integrals, and obtain the
solutions:
\begin{eqnarray}
I_{\rm 1} & = & \frac{2}{\sqrt{3}}\frac{R}{c}\le(\frac{1}{\g^2}+\phi^2_{\rm p}\ri)^{1/2}{\rm K}_{\rm 1/3}(\xi_{\rm p}),\\
I_{\rm 2} & = & i\frac{2}{\sqrt{3}}\frac{R^2}{c^2}\le(\frac{1}{\g^2}+\phi^2_{\rm p}\ri){\rm K}_{\rm 2/3}(\xi_{\rm p}), \\
I_{\rm 3} & = & -\frac{2}{\sqrt{3}}\frac{R^3}{c^3}\le(\frac{1}{\g^2}+\phi^2_{\rm p}\ri)^{3/2}
    {\rm K}_{\rm 1/3}(\xi_{\rm p}),
\end{eqnarray}
where ${\rm K}_{\rm 1/3}$ and ${\rm K}_{\rm 2/3}$ are modified Bessel functions.
Substituting the solutions of integrals into Eqs.~(17) and~(18), we get
\begin{eqnarray}
E_{\rm \pa p}(\o) & = & \frac{q\b^2\o R}{\sqrt{3\pi}c^2S_{\rm o}} {\rm e}^{i\o S_{\rm o}/c}
\le(\frac{1}{\g^2}+\phi^2_{\rm p}\ri) {\rm K}_{\rm 2/3}(\xi_{\rm p}), 
\end{eqnarray}
\begin{eqnarray}
&&\!\!\!\!\! E_{\rm \pe p}(\o) = \sqrt{\frac{2}{3\pi}} \frac{q\b\o R}{c^2S_{\rm o}} {\rm e}^{i\o S_{\rm o}/c}\!
\bd[\frac{\b}{\sqrt{2}}\le(\frac{1}{\g^2}+\phi^2_{\rm p}\ri)\! {\rm K}_{\rm 2/3}(\xi_{\rm p})
\nonumber\\
&&- i\phi_{\rm p}\le(\frac{1}{\g^2}+\phi^2_{\rm p}\ri)^{1/2} 
\le\{1+\frac{\b^2}{2}\le(\frac{1}{\g^2}+\phi^2_{\rm p}\ri)\ri\}{\rm K}_{\rm 1/3}(\xi_{\rm p}) \bd],
\nonumber\\ &&
\end{eqnarray}
where we have introduced a suffix `p' on $E_\pa$ and $E_\pe$ to indicate that they are due to a positron.

\subsection{Radiation field due to an electron}

   For an electron, the position and velocity can be taken as
\begin{equation}\vec r(t)=R\le(\sin\th,\, -\frac{\cos\th}{\sqrt{2}},\, \frac{\cos\th}{\sqrt{2}}\ri),
\end{equation}
\begin{equation}\vec \b(t)=\b\le(\cos\th,\, \frac{\sin\th}{\sqrt{2}},\, -\frac{\sin\th}{\sqrt{2}}\ri).
\end{equation}
We choose the propagation vector for electron field as
\begin{equation}\hat k=(\cos\phi_{\rm e},\, \sin\phi_{\rm e},\, 0)
\end{equation}
such that it is parallel to the $\hat k$ in Eq.~(12) and
observer receives the radiation from both particles (positron--electron). 
Here $\phi_{\rm e}$ is the angle between $\hat k$ and the electron velocity.
Using Eqs.~(30) and (31), the vector part of the integrand in Eq.~(9) 
can be written as
\begin{eqnarray}
&&\!\!\!\!\!\!\hat{k}\times(\hat{k}\times \vec \b)=\b \!\!
\left [\frac{\sin\th}{\sqrt{2}}
\hat\eps_\pa+\!\!\le(\sin\phi_{\rm e}\cos\th-\frac{\cos\phi_{\rm e}\sin\th}{\sqrt{2}}\ri) \hat\eps_\pe\!\ri]
\nonumber\\
& & 
\end{eqnarray}
where $\hat \eps_\pa=\hat z$ and $\hat \eps_\pe=- \sin\phi_{\rm e} \hat x +
\cos\phi_{\rm e} \hat y =\hat \eps_\pa\times\hat k.$
\par
The argument of the exponential in Eq.~(9) is
\begin{eqnarray}&&\!\!\!\!\!\!
\o\!\left(t-\frac{\hat k.\vec r}{c}\right)\!=\o\!\left[t-\frac{R}{c}\left(
\cos\phi_{\rm e}\sin\th-\frac{\sin\phi_{\rm e}\cos\th}{\sqrt{2}}\right)\!\right]. \nonumber\\
 && 
\end{eqnarray}
Using the identical approximations and solutions of integrals used in the previous section for
a positron field, we find the components of an electron field:
\begin{eqnarray}
E_{\rm \pa e}(\o) & = & \frac{1}{\sqrt{3\pi}} \frac{q\b^2\o R}{c^2S_{\rm o}}
{\rm e}^{i\o S_{\rm o}/c} \le(\frac{1}{\g^2}+\phi^2_{\rm e}\ri) {\rm K}_{\rm 2/3}(\xi_{\rm e}), 
\end{eqnarray}
\begin{eqnarray}
&&\!\!\!\!\! E_{\rm \pe e}(\o) = -\sqrt{\frac{2}{3\pi}} \frac{q\b\o R}{c^2S_{\rm o}} {\rm e}^{i\o S_{\rm o}/c}\!
\bd[\frac{\b}{\sqrt{2}}\!\le(\frac{1}{\g^2}\! +\phi^2_{\rm e}\ri)\! {\rm K}_{\rm 2/3}(\xi_{\rm e})\! 
\nonumber\\
&&+ i\phi_{\rm e}\le(\frac{1}{\g^2}+\phi^2_{\rm e}\ri)^{1/2}
\le\{1+\frac{\b^2}{2}\le(\frac{1}{\g^2}+\phi^2_{\rm e}\ri)\ri\}{\rm K}_{\rm 1/3}(\xi_{\rm e}) \bd].
\nonumber\\
&& 
\end{eqnarray}

\section{Polarization of radiation emitted by positrons and electrons}

The electric fields $\vec E_{\rm p}(\o)$ and $\vec E_{\rm e}(\o)$ derived in Sect.~3
describe the polarization properties of OPM. It is clear that one mode is emitted by 
positrons and other by electrons. If the two radiation fields do not bear any phase 
relation then they are expected to be incoherently superposed at the observation point.
On the other hand, if there is a phase relation then they are coherently
superposed. From the observational point of view both the cases are important, and 
we discuss them separately in the following two subsections.

\subsection{Incoherent superposition of radiation fields}

    Consider a region in the magnetosphere containing a large number of radiating
positrons and electrons.
Let each particle emit a pulse of radiation with electric field $\vec E_{\rm o}(t).$ An observer will
detect a series  of such pulses, all with same shape but random arrival times 
$t_{\rm 1},$ $t_{\rm 2},$ $t_{\rm 3},$ ..... $t_{\rm N}.$ Then the measured electric field will be
(Rybicki \& Lightman 1979)
\begin{eqnarray}
\vec E(t)=\sum_{\rm j=1}^{N}\vec E_{\rm o}(t-t_{\rm j}),
\end{eqnarray}
where N is the number of particles. Taking the Fourier transform, we find
\begin{eqnarray}
\vec E(\o)=\frac{1}{\sqrt{2\pi}}\sum_{\rm j=1}^N\int_{-\infty}^{+\infty} \vec E_{\rm o}(t-t_{\rm j})
{\rm e}^{i\o t}dt.
\end{eqnarray}
Let $u=t-t_{\rm j}$ then $du=dt.$ Therefore, we get
\begin{eqnarray}
\vec E(\o)=\vec E_{\rm o}(\o) \sum_{\rm j=1}^N {\rm e}^{i\o t_{\rm j}}.
\end{eqnarray}
\par 
   Let $dW$, $dA$ and $d\o$ be the differential incriments in energy, area and frequency,
respectively. Then the measured spectrum is given by
\begin{eqnarray}
\frac{dW}{dA d\o}&=&c\vert\vec E(\o)\vert^2 \nonumber\\
& = & c \vert\vec E_{\rm o}(\o)\vert^2 
\sum_{\rm j=1}^N \sum_{\rm k=1}^N {\rm e}^{i\o (t_{\rm j}-t_{\rm k})} \nonumber\\
& = & c \vert\vec E_{\rm o}(\o)\vert^2 \left[ N+\sum_{\rm j\neq k}^N\cos\o(t_{\rm j}-t_{\rm k})\right].
\end{eqnarray}
Since $t_{\rm j}$ and $t_{\rm k}$ are randomly distributed in the case of radiation fields which do not
have any phase relations, the second term averages to zero. Therefore, we have
\begin{eqnarray}
    \frac{dW}{dA d\o}=c\vert\vec E(\o)\vert^2 = N c \vert\vec E_{\rm o}(\o)\vert^2.
\end{eqnarray}
Hence in an incoherent superposition of radiation fields the measured intensity will be
simply a sum of intensities radiated by the individual charges. 

\subsection{Coherent superposition of radiation fields}

     The very high brightness temperature ($10^{25}$--$10^{30}$~K) of pulsars, lead to the
conclusion that the radiation must be coherent. Pacini \& Rees (1970), and Sturrock (1971)
among others were quick to point out that the observed coherence may be due to 
bunching of particles in the emission region of magnetosphere. However, the topic of 
bunching mechanism continues to be an outstanding challenge. 
If the bunches of plasma particles with sizes much smaller than a
wavelength exist then the arrival times $t_{\rm j}\approx 0,$ because
all the pulses will have a same arrival time
to order (size of bunch)/(wavelength). Then Eq.~(38) gives
\begin{eqnarray}
\vec E(\o)\approx N\vec E_{\rm o}(\o).
\end{eqnarray}
Hence the total radiation field due to a bunch of particles is equal to the vector
sum of the fields radiated by each charge.
\par
	Now, the measured spectrum is given by
\begin{eqnarray}
\frac{dW}{dA d\o}=c\vert\vec E(\o)\vert^2= N^2 c \vert\vec E_{\rm o}(\o)\vert^2.
\end{eqnarray}
Hence the coherent sum of radiation fields of a bunch of particles leads to the 
intensity, which is equal to $N^2$ times the intensities due to the individual charges.
\par
   Since the pair creation, breakdown of polar gaps and sparks are not steady state
processes, the plasma in the bunches can be neutral or nonneutral at an arbitrary time.
If the bunches consist of prominently positrons (electrons) then the 
radiation field will be a
coherent sum of $\vec E_{\rm p}$ of each positron ($\vec E_{\rm e}$ of each electron).
In such cases, the two radiation fields are orthogonally polarized, and the modes are 
said to exist disjointly. On the other hand, if bunchs are neutral then both
charges can contribute
equally and simultaneously to the radiation field. Since both the modes are coherently 
superposed in this case, we call this case joint. 
On the basis of individual pulse polarization Stinebring 
et~al. (1984a) have concluded that OPM are superposed. If OPM exist
disjointly their polarization properties are well defined, but not when they
exist jointly. In the following two subsections we consider these cases in detail.

\subsection{Presence of OPM disjointly}

	The OPM are said to exist disjointly when only one mode exists or one mode has become very
strong compared to other mode. In all such cases the polarization state is solely determined by the stronger mode. The polarization state of OPM can be  
described more accurately using the Stokes parameters:
\begin{eqnarray}
I_{\rm i} & = & E_{\rm \pa i} E_{\rm \pa i}^*+E_{\rm \pe i} E_{\rm \pe i}^* \nonumber\\
    & = & \a^2 \bd[\b^2\le(\frac{1}{\g^2}+\phi^2_{\rm i}\ri)^2 {\rm K}_{\rm 2/3}^2(\xi_{\rm i})+ 
  \phi_{\rm i}^2\bc(\frac{1}{\g^2}+\phi^2_{\rm i}\bc) \nonumber\\
&& \times \bc\{1+\frac{\b^2}{2}\bc(\frac{1}{\g^2}+\phi^2_{\rm i}\bc)\bc\}^2
{\rm K}_{\rm 1/3}^2(\xi_{\rm i}) \bd], 
\end{eqnarray}
\begin{eqnarray}
Q_{\rm i} & = & E_{\rm \pa i} E_{\rm \pa i}^*-E_{\rm \pe i} E_{\rm \pe i}^* \nonumber\\
 & =& -\a^2\phi_{\rm i}^2\le(\frac{1}{\g^2}+\phi^2_{\rm i}\ri)\!\! \le\{1+\frac{\b^2}{2}\le(\frac{1}{\g^2}+\phi^2_{\rm i}\ri)\ri\}^2\!{\rm K}_{\rm 1/3}^2(\xi_{\rm i}), 
\nonumber\\ &&
\end{eqnarray}
\begin{eqnarray}
U_{\rm i} & = & 2 {\rm Re}[E_{\rm \pa i}^* E_{\rm \pe i}]\nonumber \\
    & = & \eta \a^2\b^2 \le(\frac{1}{\g^2}+\phi^2_{\rm i}\ri)^2 {\rm K}_{\rm 2/3}^2(\xi_{\rm i}), 
\end{eqnarray}
\begin{eqnarray}
V_{\rm i} & = & 2 {\rm Im}[E_{\rm \pa i}^* E_{\rm \pe i}]\nonumber\\
    & = & -\sqrt{2}\a^2\b \phi_{\rm i}\le(\frac{1}{\g^2}+\phi^2_{\rm i}\ri)^{3/2} \le\{1+\frac{\b^2}{2}\le(\frac{1}{\g^2}+\phi^2_{\rm i}\ri)\ri\} \nonumber\\ 
&&\times {\rm K}_{\rm 1/3}(\xi_{\rm i}) {\rm K}_{\rm 2/3}(\xi_{\rm i}),
\end{eqnarray}
where 
\begin{eqnarray}\xi_{\rm i}&=&\frac{\o R}{3c}\le(\frac{1}{\g^2}+\phi^2_{\rm i}\ri)^{3/2}, \quad\quad \a=\sqrt{\frac{2}{3\pi}}\frac{q\b\o R}{c^2 S_{\rm o}},
\nonumber
\end{eqnarray}
i $=$ p and $\eta=+1$ for positrons, and i $=$ e and $\eta=-1$ for electrons.
The intensity and polarization angle of the linearly polarized radiation are:
\begin{equation}
L_{\rm i}=\sqrt{U_{\rm i}^2+Q_{\rm i}^2}, 
\end{equation}
\begin{equation}
\psi_{\rm i}=\frac{1}{2} \tan^{-1}\le(\frac{U_{\rm i}}{Q_{\rm i}}\ri).
\end{equation}

Due to the action of forces $\vec F_{\rm Bi}$ and $\vec F_{\rm ci},$ particle orbital planes become inclined with respect to the plane of
magnetic field line. The radiation beams  of the two charges appear on either sides of 
the plane of magnetic field line.
Therefore, we define $\phi_{\rm p}=\phi-\frac{1}{\g}$ and $\phi_{\rm e}=\phi+\frac{1}{\g},$ where $\phi$
is the angle between the plane of magnetic field line and $\hat k.$ 
Using $\g=300,$ $\nu=1$~GHz and $R=10^6$~cm, we computed the polarization parameters 
$I_{\rm i},$ $L_{\rm i},$ $V_{\rm i}$ and $\psi_{\rm i},$ and plotted as functions of $\phi$ 
in Fig.~4. The continuous line curves describe the positron radiation field while the broken line curves for an electron radiation field. 
About the particle orbital planes, i.e., about $\phi_{\rm p}=0$ and $\phi_{\rm e}=0,$ intensity and linear polarization 
have maxima while circular polarization undergoes sense reversal. Therefore, when observed with
line-of-sight lying in the range $-\frac{1}{\g}<\phi<\frac{1}{\g}$ as indicated by a two way arrow in 
Fig.~4(c), one mode becomes right hand circularly polarized 
($V_{\rm p}>0)$ while the other becomes left hand circularly polarized 
($V_{\rm e}<0).$ Individual pulses from PSR~B2020+28 (Cordes et~al. 1978)
and PSR~B0950+08 (Fig.~10g) indicate OPM tend to have opposite circular
polarization. Figure 4(d) shows the polarization angle as a function of 
$\phi,$  clearly, two modes are 
orthogonally polarized when $I_{\rm i}$ and $L_{\rm i}$ are in maxima. Since particles are constrained to
follow the curved field lines, the polarization angle of each mode swings in accordance with
the RVM. In the case of  
PSR~B0329+54 at 408~MHz, Gil \& Lyne (1995) have clearly shown that each of these modes
is well described by the RVM. 
\par
  The Stokes parameters of $\vec E_{\rm p}(\o)$ and $\vec E_{\rm e}(\o)$ clearly
indicate OPM are elliptically polarized. 
The radiation field $\vec E_{\rm p}$ is polarized with $\sim 45^o$ while $\vec E_{\rm e}$
with $\sim -45^o$ with respect to the plane of magnetic field line.
Figure~5 illustrates the polarization ellipses of OPM in the coordinate system with axes along the
unit vectors $\hat \eps_\pa,$ $\hat \eps_\pe$ and $\hat k,$ where the unit vector $\hat \eps_\pa$ is
parallel to the radius of curvature ($\vec \rho$) of magnetic field line. 

\begin{figure}
\vskip -3.0 truecm 
\epsfxsize=16. truecm
\epsfysize=26.0 truecm
\centerline{\hskip -1.0 truecm\epsffile[0 0 596 842]{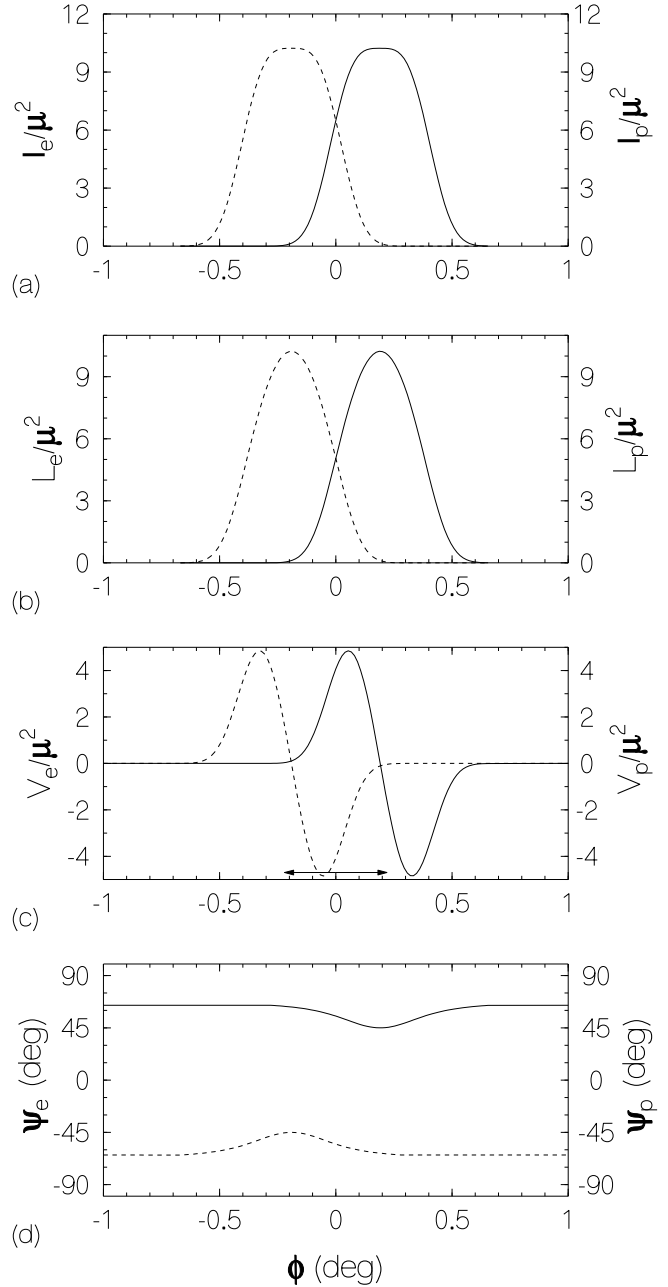}}
\vskip -3.0 truecm
\caption{Polarization parameters versus $\phi$ for the radiation due 
to a positron (continuous line curves) and an electron (broken line curves). The labels on
vertical axes of panels (a), (b) and (c) are dimensionless as we have normalized with 
$\mu^2=[q\b^2\o R/(3\pi)^{1/2}c^2S_{\rm o}\g^2]^2.$ }
\end{figure}

   As mentioned in Sect.~4.2, particle generation processes such as direct emission
of electrons from the stellar surface due to the building up of high potential difference at the
polar cap and the pair creation are not steady processes. They operate on time scales less than microseconds,
therefore, the plasma which flows along the field lines is not always neutral. On time scales such as
sampling time, it may look either negative or positive. In all such cases, the observer receives OPM
at $\sim\pm 45^o,$ depending upon the sign of plasma which exist at that particular instant. One mode is not radiated even when the plasma is neutral, if the energy of one kind of charge 
happens to fall below the threshold to radiate.

\begin{figure}
\vskip -11.0 truecm \epsfysize=19. truecm
\centerline{\hskip 6.0 truecm\epsffile[0 0 596 842]{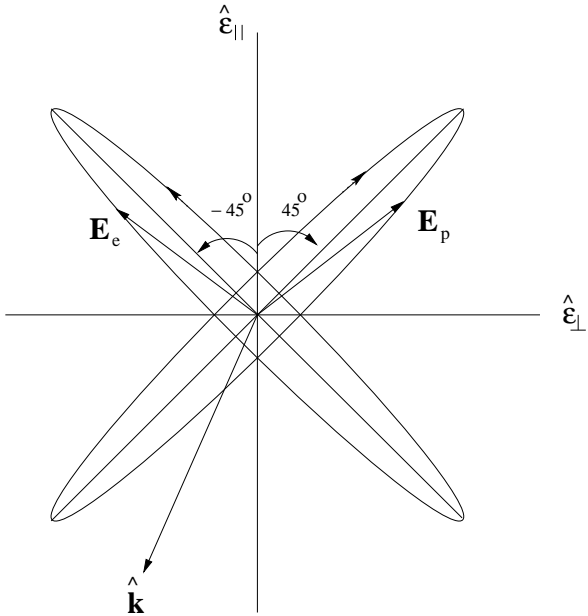}}
\caption{Representation of OPM in a coordinate system with axes parallel to 
the orthogonal unit vectors $\hat \eps_\pa,$ $\hat\eps_\pe$ and $\hat k,$ where 
$\hat \eps_\pa$ is parallel to the radius of curvature $({\bf\rho})$ of magnetic field line.}
\end{figure}

\subsection{Presence of OPM jointly}

  If the charges are close enough to each other (e.g. bunch), we may expect the fluctuations in 
amplitudes and phases of radiation fields will not be independent at the observation point. In the
limit where the distance between the two charges is much less than a wavelength, the amplitudes 
vary in phase (Born \& Wolf 1986). Therefore, when the Fourier components $\vec E_{\rm p}(\o)$ and $\vec E_{\rm e}(\o)$ 
are coherently superposed, we get
\begin{equation}
\vec E(\o)=\vec E_{\rm p}(\o)+\vec E_{\rm e}(\o).
\end{equation}
\par
The Stokes parameters of total field can be defined similar to those
(Eqs.~43--46) for OPM. Again using $\g=300,$ $\nu=1$~GHz and $R=10^6$~cm,
we computed polarization parameters $I,$ $L,$ $V$ and $\psi$ of 
$\vec E(\o),$ and plotted as functions of $\phi$ in Fig.~6. 
The two humps at $\phi\sim 0.2^o$ and $-0.2^o$ in Figs. 6(a) and (b) are due to the
dominance of emission from positrons and electrons at those angles, respectively. 
Figure~6(c) shows the sense reversal of circular polarization about $\phi=0,$ i.e.,
about the plane of magnetic field line. The polarization angle of $\vec E(\o)$ is plotted in Fig.~6(d), and the
swing arises due to the coherent superposition of OPM. 
This type of polarization angle swings are quite evident in 
micropulses and subpulses.
\par
  Consider a coordinate system-xyz such that the major axes of polarization ellipses of $\vec E_{\rm p}(t)$ 
and $\vec E_{\rm e}(t)$ are parallel to the axes x and y, respectively, as shown in Fig.~7. For the sake of illustration of OPM we represent them as 
\begin{equation}
\vec E_{\rm p}(t) = E_{\rm xp}\cos(\o t+\d_{\rm p})\hat x+E_{\rm yp}\sin(\o t+\d_{\rm p})\hat{y},
\end{equation}
\begin{equation}
\vec E_{\rm e}(t) = E_{\rm xe} \sin(\o t+\d_{\rm e})\hat x+E_{\rm ye}\cos(\o t+\d_{\rm e})\hat{y},
\end{equation}
where $E_{\rm xi},$ $E_{\rm yi}$ and $\d_{\rm i}$ are the amplitudes and initial 
phases. The period of rotation of 
electric fields is $T=1/\nu.$ For $\nu=1$~GHz we find $T=10^{-9}$~s,
much smaller than the sampling interval in a typical observation 
$(\geq 1\mu$s).
\par
Let $\d=\d_{\rm p}-\d_{\rm e}$ be the phase difference between OPM, and $\psi$ be the polarization angle of total field
$\vec E(t)$ measured counter-clockwise from the x-axis. In Fig.~7, the ellipse drawn with dots indicate the 
polarization state of $\vec E(t)$ with the sense of rotation can be clockwise or counterclockwise or even 
nil, depending upon strengths of circular polarization of OPM.
For the values of $\d$ in the range $-\frac{\pi}{2}$ to $\frac{\pi}{2}$ we find 
$\psi$  lies between $0$ and $\frac{\pi}{2}.$ On the other hand for values of $\d$ in the range
$\frac{\pi}{2}$ to $\frac{3\pi}{2},$ $\psi$ takes values between $\frac{\pi}{2}$ and $\pi.$ 
The polarization state of $\vec E(t)$ follows closely the stronger mode.
\par
It is probable that separate streams of positrons and electrons exist in the pulsar
magnetosphere (Sturrock, 1971). The mode changing observed in PSR~B0329+54 by Hesse et~al.
(1973) indicate the channeling of particles along field lines is not steady. Assume that there
exist two separate and closely spaced streams: one
consists of mainly positrons and other electrons at longitudes 
$\phi_1$ and $\phi_2,$ respectively.  When the observer's line-of-sight moves from $\phi_1$ to $\phi_2$ there
will be an orthogonal {\it jump} in the polarization angle.  Now, what happens to degree of polarization
during those jumps? Let $\d t$ be the temporal separation between $\phi_1$ and $\phi_2.$ When one
approaches $\phi_2$ starting from $\phi_1,$ field $\vec E_{\rm p}$ becomes
progressively weaker while $\vec E_{\rm e}$
gets stronger. To find the polarization angle distribution during $\d t$, let us divide the interval $\d t$ into $n$ subintervals having the widths equal to 
the period $(T=1/\nu)$ of rotation  of $\vec E(t).$ If $\d$ lies between $-\frac{\pi}{2}$ 
and $\frac{\pi}{2}$ then $\vec E(t)$ will have ellipses $t_1=T$, 
$t_2=2T$, .... $t_n=nT=\d t$ with polarization angles $\psi_1,$ $\psi_2,$ .... $\psi_n,$ respectively, as shown in Fig.~8. On the
other hand when $\d$ lies between $\frac{\pi}{2}$  and $\frac{3\pi}{2}$ we get the ellipses $\tau_1=T$, 
$\tau_2=2T$, .... $\tau_n=nT=\d t$ with polarization angles $-\psi_1,$ $-\psi_2,$ .... $-\psi_n.$
If the observations are made with the time resolution, which is of the order of $\d t,$

\begin{figure}
\vskip -3.0 truecm 
\epsfxsize=16. truecm
\epsfysize=25.5 truecm
\centerline{\hskip 0.0 truecm\epsffile[0 0 596 842]{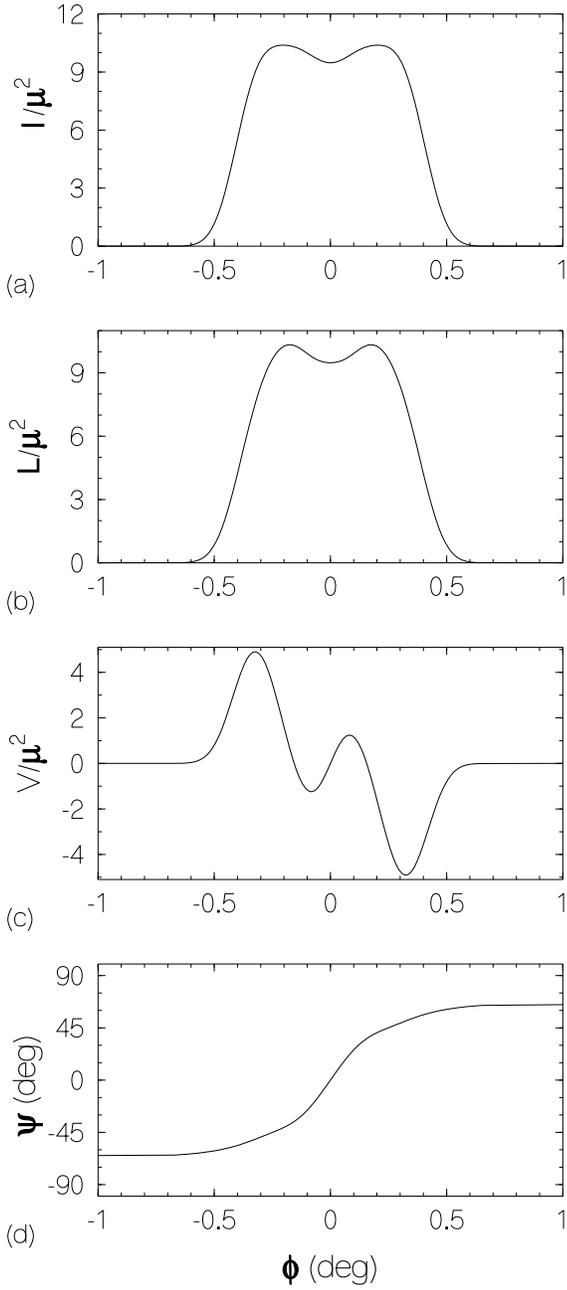}}
\vskip -2.5 truecm
\caption{Polarization parameters versus $\phi$ for the total radiation field 
${\bf E}(\o).$ The labels on vertical axes of panels (a), (b) and (c) are 
dimensionless as we have
 normalized with $\mu^2=[q\b^2\o R/(3\pi)^{1/2}c^2S_{\rm o}\g^2]^2.$ }
\end{figure}

\begin{figure}
\vskip -9.7 truecm \epsfysize=18. truecm
\hskip 3.0 truecm\epsffile[0 0 596 842]{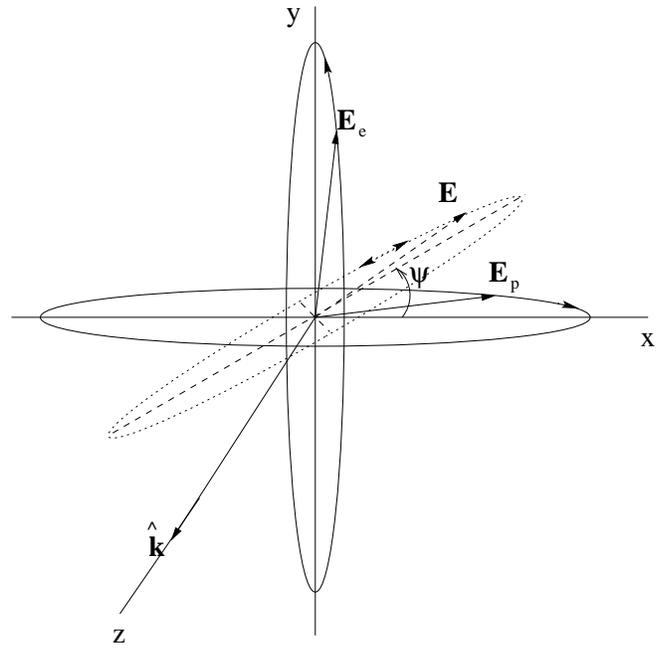}
\vskip 0.0 truecm
\caption{Representation of OPM and total electric field ${\bf E}(t)$ in the coordinate system--xyz.
The angle $\psi$ represent the polarization angle of total field.}
\end{figure}

\begin{figure}
\vskip -10.5 truecm \epsfysize=19.0 truecm
\hskip 3.0 truecm\epsffile[0 0 596 842]{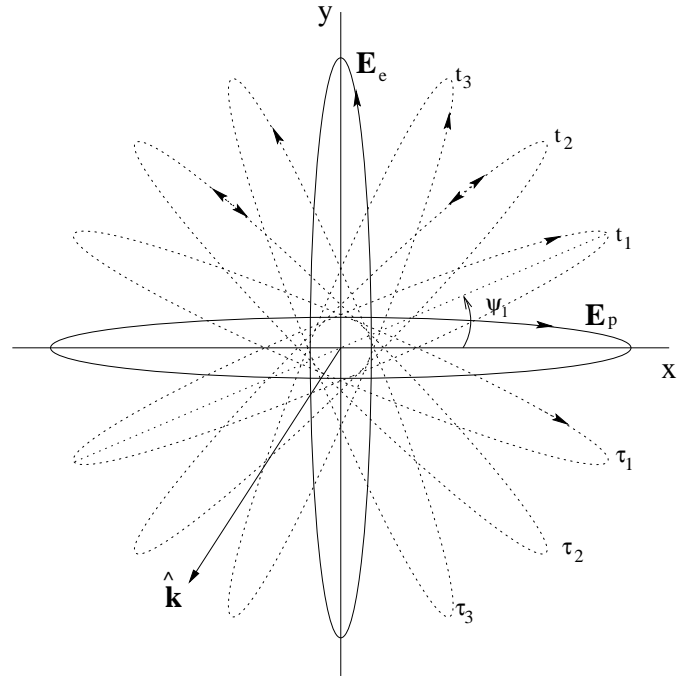}
\caption{Orthogonal polarization angle jumps occur when one mode say ${\bf E}_{\rm p}$ becomes
weaker while the other mode ${\bf E}_{\rm e}$ gets stronger or vice versa over a small interval of pulse phase during pulsar rotation.}
\end{figure}

\noindent
we get total depolarization. This is because during the integration the electric fields oriented in 
all directions (doted line ellipses) are superposed. The general result of observation that the 
orthogonal polarization angle jumps are accompanied by percentage of polarization going to zero, 
are caused by such an effect. For example, see the pulses of PSR~B1804$-$08 and PSR~B1905$+$39 
by Xilouris et al. (1991).

\section{Discussion with an application to polarization of PSR~B0950+08}

    In the previous sections we discussed the polarization properties of radiation emitted by 
positrons and electrons while moving along a curved magnetic field line. Here we 
extend our discussion to
the emission from plasma streaming along many field lines. In
the curved region of magnetic field lines, for the reasons given in
Sect.~2, positrons move to one side of field line while 
electrons to other side as shown in Fig.~9.
Let us consider three observing positions: A at the left side of the pulse, B at the middle and 
C at the right side. 

\begin{figure}
\vskip -1.0 truecm
\epsfysize=11.0 truecm
\centerline{\hskip -1.0 truecm\epsffile[18 176 515 673]{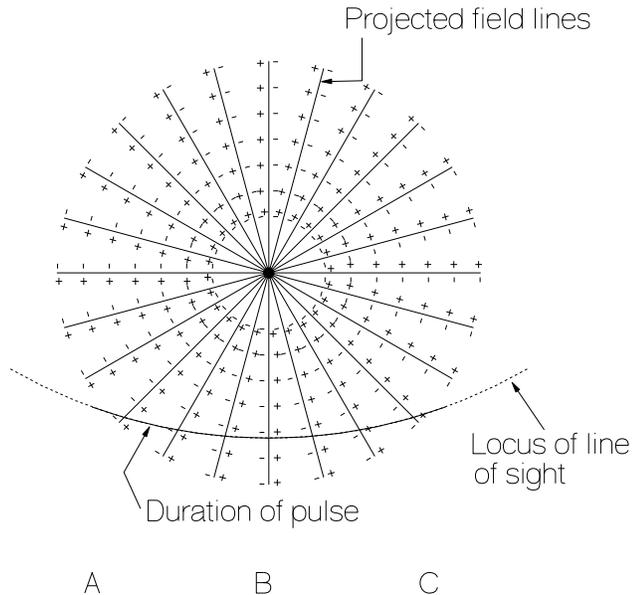}}
\vskip -2.0 truecm
\caption{Projected field lines loaded with positrons ($+$ signs) and 
electrons ($-$ signs). Points A, B and C are the three observing positions.}
\end{figure}

At the position A,
observer mainly receives radiation from the left side of planes of
magnetic field lines, and the circular polarization of OPM tend show
opposite senses (see Fig.~4c).
An observer at position B (fiducial plane) views 
particle trajectory planes edge-on, therefore, the circular polarization of 
each mode undergoes sense reversal. Finally, an observer at C receives 
radiation mainly from the right side of planes of magnetic field lines, and
circular polarization of each mode becomes opposite to its
value in position A. If the OPM exist disjointly then
the polarization angle of each mode swings in accordance with the RVM
when the observers line--of--sight moves from point A to C.
\par
    Radhakrishnan \& Rankin (1990) have identified the two extreme types of circular polarization
signatures: an antisymmetric type (Fig.~6d) wherein the circular polarization changes sense at the 
middle of the pulse, and a symmetric type wherein it is predominantly of one sense. Our model can easily
explain the antisymmetric type of circular polarization, while for symmetric type we need
further analysis on the high resolution data of individual pulses.

\subsection{Observations of PSR~B0950+08}

 The  pulsar PSR~B0950+08 was observed in 1994 April by using the 100--m Effelsberg radiotelescope. 
Using a tunable HEMT--receiver with a system temperature of $28$~K, observations
were made at the center frequency of 1.71~GHz with a bandwidth of 40~MHz.
The two circular polarizations are separated in the receiver and amplified. The signal is then
fed into an adding polarimeter, a passive device with four output channels which allows further
online signal processing.
The pulse--smearing caused by the dispersion due to the interstellar medium is than removed
using an online dedispersion device. This is a four unit $60\times667$~kHz filterbank.
The output of each channel is then detected and converted into a digital signal by a
fast A/D converter. After a time delay according to
the dispersion measure, the outputs of all channels are added and than recorded by the backend.
After a careful calibration procedure, Stokes parameters are obtained from the
four recorded output channels. A detailed system description and
the calibration procedure are given by von~Hoensbroech \& Xilouris 
(1997).
At frequency 1.71~GHz, we recorded about 1200 pulses with a time resolution of 0.24~ms.
\par
We consider only the main pulse as the interpulse is too weak to reproduce the properties of OPM.
The average polarization parameters are plotted as functions of pulse phase $\phi$ in 
Fig.~10(a). The continuous line curve indicates intensity $(I)$ variation while broken and
doted ones indicate linear $(L)$ and circular $(V)$ polarization, respectively, in arbitrary units.
\par
     The gray-plots show the frequency of occurrence of OPM at different pulse phases, and have become 
powerful tools in analyzing the pulsar polarization properties.
The darkest regions represent the most probable regions of occurrence.
For each pulse phase bin where $I$ is above 
$3\sigma$ level of its value in the off pulse region, gray-plots were computed 
for $L$, $V$, $I$ and polarization angle $\psi.$
All those phase bins, where the condition $L^2+V^2\leq I^2$ was
not met, were excluded as they lead to spurious interpretation of

\begin{figure*}
\vskip -2.0 truecm \epsfysize=24.5 truecm
\centerline{\hskip 1.5 truecm\epsffile[0 0 596 842]{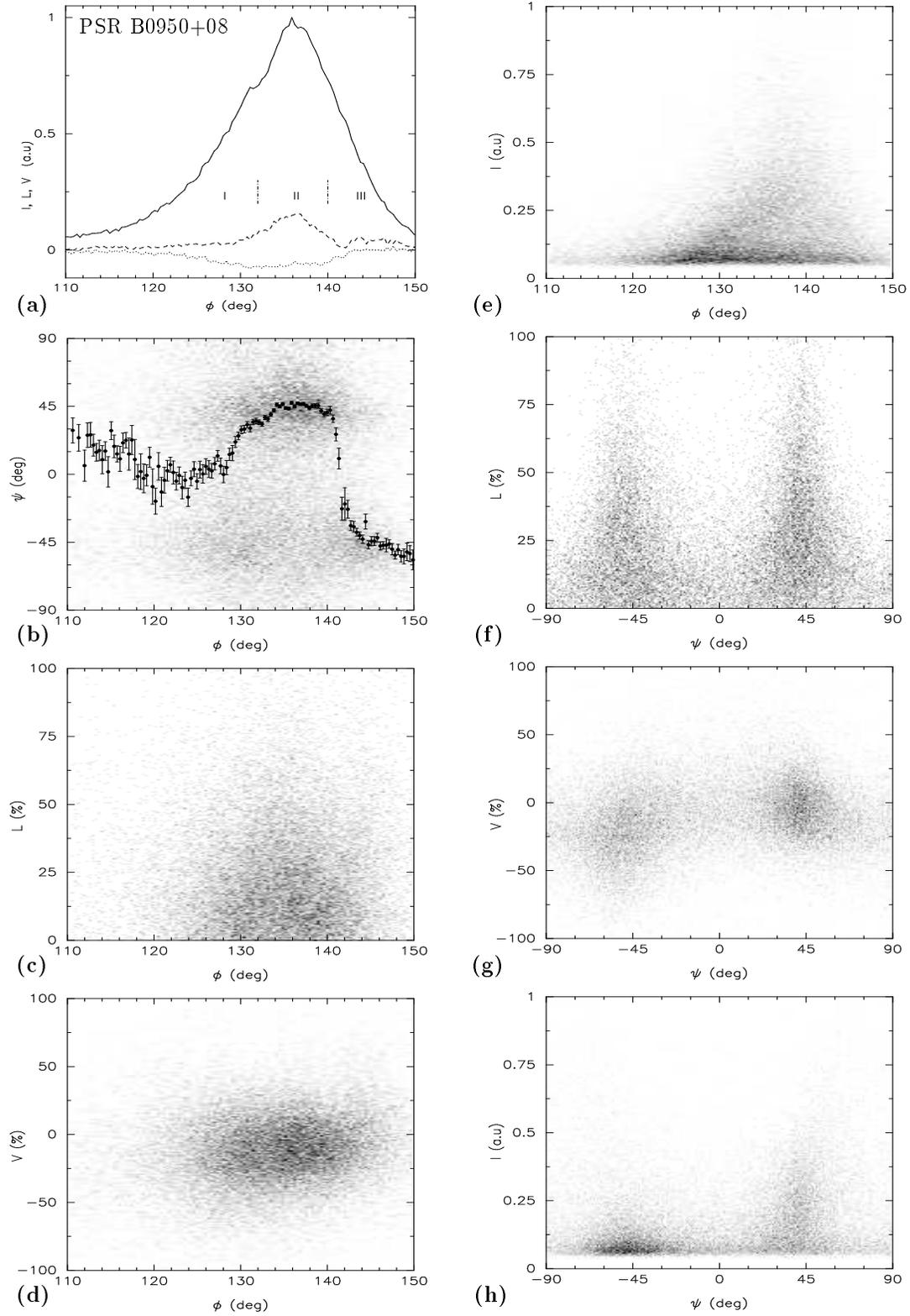}}
\vskip -1.0 truecm
\caption{Polarization histograms of pulsar PSR~B0950+08 at frequency 1.71~GHz:
(a) average pulse with arbitrary units (a.u), (b)~polarization angle $\psi$, (c) \& (f) percentage of 
linear polarization,
(d) \& (g) percentage of circular polarization, and (e) \& (h) intensity. The angle $\phi$ denotes the
 pulse phase.}
\end{figure*}

\noindent
polarization.  The points with error bars in Fig.~10(b) indicate the integrated polarization angle 
superposed over the polarization angle gray-plot. Figures 10(c)--(e) represent the
gray-scale maps of percentage
of linear polarization $[L(\%)=100L/I],$ circular polarization $[V(\%)=100V/I]$ and $I,$
respectively, while Figs.~10(f)--(h) represent the gray-scale scatter plots of $L(\%),$ $V(\%)$ and $I$ versus the
polarization angle $\psi .$
\par
    The polarization angle gray-plot (Fig.~10b) shows the two most preferred tracks close to $45^o$ and 
$-45^o$ position angles in the $\psi$ versus $\phi$ plane, in agreement with our model (Fig.~4d). 
The frequency of occurrence of OPM with respect to the polarization angle is shown
in Fig.~11, proving the importance of OPM in the pulsar radiation. 
At any pulse longitude the average polarization angle curve follows closely the mode which is more intense, as 
indicated by Fig.~10(b). The uniform or random component of polarization angle  
is probably due to the coherent superposition of OPM (see Figs. 6d\&7). 
This idea is also supported by the observation that the random component
becomes significant only at those pulse longitudes where
both modes exit.  For example, see the polarization angle 
displays of PSR~B0823+26, B0950+08 (Stinebring et al. 1984a), B0834+06
(Stinebring et al. 1984b), and B0329+54 (Gil \& Lyne 1995).  McKinnon
and Stinebring (1996) have also suggested that the 
random component may arise from superposed modes. Barnard \& Arons (1986) have proposed $\o^{-2}$
frequency dependence for angular separation of OPM, but our model do not predict any 
appreciable frequency dependence.

\begin{figure}
\vskip -3.0 truecm \epsfysize=10.0 truecm\epsfxsize=9.0 truecm
\centerline{\hskip 0.5 truecm\epsffile[25 18 587 774]{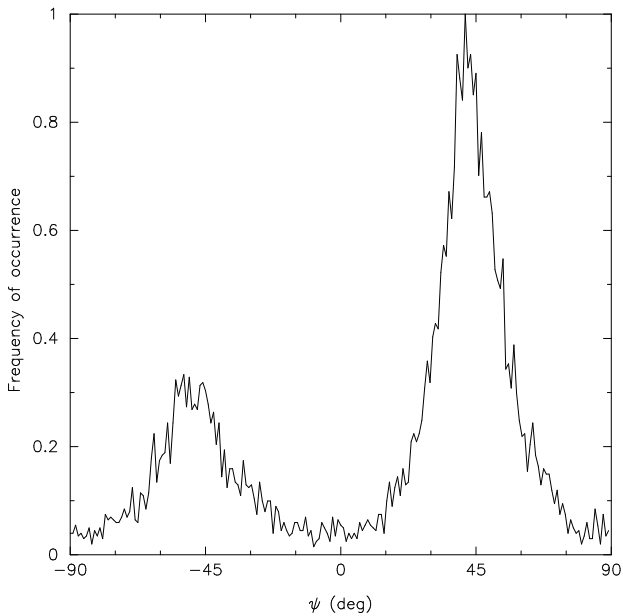}}
\caption{Frequency of occurrence of pulses with respect to the
polarization angle.}
\end{figure}

     On the basis of relative frequency of occurrence of OPM at different pulse longitudes as indicated by 
Figs.~10(b), we may identify pulse longitude ranges I, II and III, as marked in Fig.~10(a).
It is clear that both modes exist at all pulse longitudes but one mode dominates over the other at some
longitudes. For example, in region II the mode (say mode~1) with $\psi\sim 45^o$
dominates while in region III the other mode (mode~2) with $\psi\sim -45^o$ dominates. 

\begin{figure}
\vskip -4.0 truecm
\epsfysize=21.0 truecm
\centerline{\hskip 0.0 truecm\epsffile[0 0 596 842]{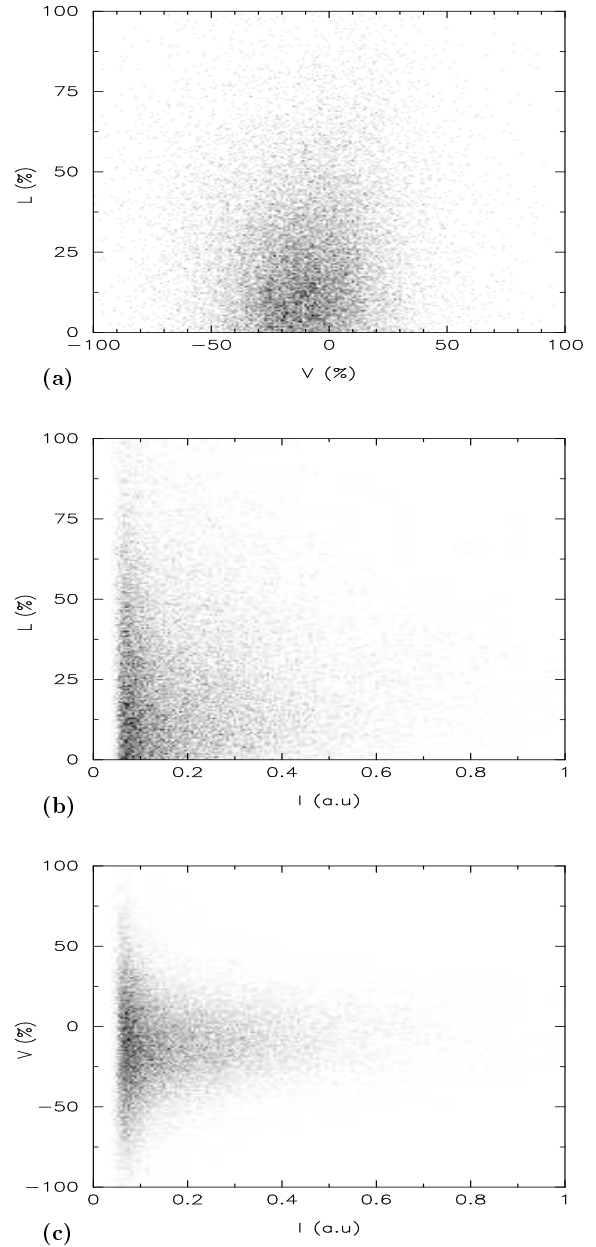}}
\caption{Grey--scale plots representing the correlations between
L(\%), V(\%) and I in the individual pulses.}
\end{figure}

    When the frequency of occurrence of OPM is very high as in region II, it is likely that more often they 
simultaneously exist. If so then they get superposed either coherently or incoherently depending upon 
whether there is any phase relation between them or not. 
Figure~10(d) shows, at any pulse longitude, circular polarization can be of either 
sense depending upon from which side of particle trajectory the radiation is received (see Fig.~4c).
   The distribution of different pulse intensities (arbitrary units) versus $\phi$ is shown in Fig.~10(e).
\par
   In revealing the properties of OPM, Figs.~10(f)--(h) are much more expressive than Figs.~10(c)--(d) as the
earlier ones directly display the distributions against the polarization angle $\psi.$ 
Figures~10(f) and (g) indicate that if the modes exist disjointly, they appear at $\sim\pm 45^o$ with high 
inear and circular polarization, but the random component is less polarized.
Figure~10(g) shows OPM tend have opposite circular polarization (see Fig.~4c\&d). Figure 10(h) shows the
distribution of intensities of OPM versus the polarization angle, and it indicates the mode~1 is stronger than the mode~2.
\par
     From the theoretical point of view, it is very important to find the correlations between 
$L(\%),$ $V(\%)$ and $I.$ Using the individual pulse data, we computed the grey-plots (Fig.~12), 
which represent the correlations between different polarization parameters.
Figure 12(a) shows {\it linear polarization becomes maximum when the circular polarization is at minimum,} 
a prediction of curvature radiation (see Figs. 4b\&c).
\par
The behaviour of $L(\%)$ with respect to $I$ is shown in
Fig.~12(b). It shows anticorrelation between $I$ and $L(\%)$ at higher intensities.
The sharp cutoffs close to the vertical axis is due to the condition that $I$
is above $3\sigma$ level. 
Manchester et~al. (1975) and Xilouris et~al. (1994) have also indicated the anticorrelation between $I$ and $L(\%)$.
  The behaviour of circular polarization with respect to the intensity is shown in Fig.~12(c). At higher
intensities, circular polarization is lower. The reason could
be the superposition of OPM with opposite senses of circular polarization.

\section{Conclusion}

   The motion of particles along the curved field lines cannot be described in analogy with the motion of
particles in a central force. The radiation emitted by positrons and electrons while moving along the 
curved magnetic field lines is probably orthogonally polarized. The radiation is highly polarized 
when the OPM are not superposed. The polarization angle of each mode swings in accordance with the 
RVM. However, the polarization angle swings observed in micropulses and subpulses
are often found to be contrary to the predictions of RVM. 
We expect such swings are produced when the OPM are coherently superposed.
Our model do not predict any appreciable frequency dependence on
the angular separation between OPM. The circular polarization of OPM tend to have
opposite senses.
\par
  The observations of PSR~B0950+08, particularly, polarization angle, linear and circular
polarizations of OPM are in agreement with our model. The polarization histograms clearly indicate that
the depolarization is due to the superposition of OPM. This effect becomes much more severe at higher
frequencies as the coherence--length which is of the order of wavelength becomes small leading to 
an incoherent superposition of OPM.

\begin{acknowledgements}

    I would like to thank L. A. Nowakowski and K. Rozga for confirming the solutions of integrals,
and A. v. Hoensbroech and M. Kramer for their help in reducing the data.
It is a pleasure to thank D. Lorimer for his comments on the manuscript, and A. Jessner,
A. G. Lyne, W. Kundt, G. Smith and R. Wielebinski for several interesting and stimulating
discussions.
This work was supported by a fellowship of the Alexander-von-Humboldt foundation.
\end{acknowledgements}

\end{document}